\documentstyle[aps,preprint,epsf,floats]{revtex}
\draft
\def\lsim{\mathrel{\rlap{\lower4pt\hbox{\hskip1pt$\sim$}}
    \raise1pt\hbox{$<$}}}         
\def\gsim{\mathrel{\rlap{\lower4pt\hbox{\hskip1pt$\sim$}}
    \raise1pt\hbox{$>$}}}         
\begin{document}
\title{Perturbative Effective Theory in an Oscillator Basis?}
\author{W. C. Haxton and T. Luu}
\address{Institute for Nuclear Theory, Box 351550, and Department
of Physics, Box 351560, \\
University of Washington, Seattle, WA 98195}
\date{\today}
\maketitle
\begin{abstract}
The effective interaction/operator problem in nuclear physics 
is believed to be highly nonperturbative, requiring 
extended high-momentum spaces for accurate solution.  We trace this
to difficulties that arise at both short and long distances 
when the included space is defined in terms of a basis of 
harmonic oscillator Slater determinants.  We show, in the
simplest case of the deuteron, that both difficulties can be
circumvented, yielding highly perturbative results in the potential even
for modest ($\sim$ 6$\hbar \omega$) included spaces.   
\end{abstract}
  
\vspace{0.5cm}

There is an extensive literature on attempts to relate
the effective interaction, needed in any description of nuclei 
based on a finite set of low-momentum basis states, to the
underlying, more singular $NN$ interaction.  As Barrett \cite{bb}
has summarized, one hope of investigators in the 1970s was
that $H^{eff}$ might be expanded perturbatively in either the
bare potential $V$ or in the $G$ matrix, the sum of all 
two-nucleon ladder diagrams for scattering in the excluded,
high-momentum space.  While some phenomenological success
was achieved by selecting certain diagrams \cite{kuo}, more systematic
treatments provided little indication of convergence.
For example, the third-order calculations of Barrett and
Kirson \cite{bk} yielded a correction of the same size but opposite in sign
to the second-order result.  At about the same time Shucan
and Weidenmuller \cite{sw} demonstrated in a toy model that strongly
coupled ``intruder'' states -- states low in energy having 
little overlap with the model space -- generically lead to 
poorly convergent expansions.  These disappointing results 
led many practioners to turn to phenomenological $H^{eff}$s --
an approach that is also lacking because it fails to provide
a basis for generating effective operators consistent with
$H^{eff}$.

The lack of a first-principles technique for effective interactions
had a discouraging effect on the field for a
number of years.  However several recent developments -- 
including the rapid growth of computing power and interest
in effective field theory \cite{beane} -- have encouraged new attempts to
calculate $H^{eff}$ in a systematic, controlled way.

One step in this direction was reported recently \cite{hs}: for an
included space consisting of a complete set of harmonic
oscillator (HO) Slater determinants of energy $\le \Lambda_{SM} \hbar \omega$,
the deuteron and $^3$He were treated as exact effective theories (ET).
The point was to illustrate crucial aspects of 
ETs that are generally absent in models like the 
shell model (SM), including nontrivial wave function normalizations,
the many-body nature of effective interactions, the rapid 
evolution of $H^{eff}$ matrix elements with changes in the
model space, and the crucial role of consistent effective
operators.

The starting point for the calculation of Ref. \cite{hs} is the Bloch-Horowitz
equation \cite{bh}.  For a low-momentum ``included space'' defined by
$\Lambda_{SM}$ and a HO size parameter $b$, $H^{eff}$ 
will be translationally invariant.  (Note $\hbar \omega = 
\hbar^2/M b^2$, with $M$ the nucleon mass.)  Defining the projection
operator onto the high-momentum Slater determinants by $Q(\Lambda_{SM},b)$, 
one finds
\begin{eqnarray}
H^{eff} &=& H + H {1 \over E - Q H} Q H \nonumber \\
H^{eff} |\psi_{SM} \rangle &=& E |\psi_{SM} \rangle ~~~~ |\psi_{SM} \rangle
= (1 - Q) |\psi \rangle
\end{eqnarray}
where $|\psi \rangle$ is the exact wave function, 
$H |\psi \rangle = E |\psi \rangle$, and $H=T+V$ is the sum of
relative kinetic energy and potential terms.  This equation has to be 
solved self-consistently for each desired but unknown eigenvalue
$E$.  The method introduced in 
\cite{hs} provided an efficient solution to the self-consistency 
problem, constructing the needed Green's function as a function
of $E$ by a method based on the Lanczos algorithm:
with this technique one iteratively extracts the high-momentum
spectral information most
relevant to the construction of the Green's function.

This approach, in principle, can be extended to heavier systems.
In practice, however, the integration over high-momentum states 
becomes more challenging with increasing $A$: in \cite{hs}, where the Argonne $v18$
potential \cite{av18} was employed, high-momentum integrations up to 
$\Lambda_\infty \sim 140$ ($\sim$ 60) were necessary to achieve
1 keV (25 keV) accuracy.  In this letter we argue that this
integration can be performed in a much simpler way.

We begin by investigating why the high momentum integration is
nonperturbative.  One can envision moving the SM scale $\Lambda_{SM}$
very close to the necessary $\Lambda_\infty$, which clearly 
makes the excluded-space contribution
to the BH equation a small correction.  As the energy denominator
in Eq. (1) is then very large, one might assume that the 
high-momentum problem is now perturbative.
  
To test this conclusion we expand the excluded-space contribution
to Eq. (1) 
\begin{equation}
H{1 \over E-QH}QH = H \left[{1 \over E-QT} + {1 \over E-QT} QV
{1 \over E-QT} + ... \right] QH
\end{equation}
The order-by-order results ( dashed lines in
Fig. 1a) are quite curious.  The total excluded-space contribution,
given the very high value of $\Lambda_{SM}$ = 70 chosen,
accounts for only 20 keV of binding energy; 85\% of this 
is generated by the first term in Eq. (2).  However 
subsequent order-by-order corrections appear to converge to
a value a few keV above the true binding energy.  Only after
$\sim$ 1000 orders of perturbation theory is the correct
binding energy slowly achieved.  (Very similar results are obtained
if one uses the HO Hamiltonian $T+V_0$, instead of $T$, as the
unperturbed Hamiltonian.)

In Figs. 1b and 1c this behavior is explored for two matrix
elements of $H^{eff}$.  Most matrix elements converge rapidly,
like the $1s-1s$ example of Fig. 1b.
The exceptions are those where the bra or ket lies in the last
included shell, i.e., $\Lambda_\alpha = \Lambda_{SM}$, such as 
the $n=36$ $s$-state case of Fig 1c).  This
associates the poor convergence with $T$ which,
because of the raising/lowering properties of the 
gradient operator in a HO basis, connects only 
such states to the excluded space.
  
It is helpful to note that a HO-basis ET differs from
effective field theories in that the expansion
is around an intermediate momentum $q_0 \sim 1/b$, rather than
$q_0 \sim 0$.  Fig. 1 shows that this expansion then induces
a familiar long-distance problem: because HO wave functions fall off too
sharply, the correct asymptotic wave function can only be 
achieved by scattering through a very large number of high-momentum
oscillator states.  Consequently a poorly 
convergent ``tail'' exists regardless of how high $\Lambda_{SM}$
is chosen: a larger $\Lambda_{SM}$ restricts the unresolved tail
to larger $r$ and thus limits its numerical significance,
but does not in principle make the problem perturbative.

The solution to this problem is not entirely trivial.  The HO
basis is essential because of the center-of-mass separability
it provides.  Because the relevant operator is the relative
kinetic energy
\begin{equation}
T = {1 \over 2} \sum_{i \neq j=1}^A T_{ij}
\end{equation}
where $T_{ij} = (\vec{p}_i-\vec{p}_j)^2/2AM$, the missing 
long-distance correlations are two-body.
As the problem is associated with $T$ hopping from the
included space at large $r$ with a large amplitude, we
rearrange the BH equation so that the Green's function
is sandwiched between $V$s, thereby cutting off large-$r$
propagation \cite{hlnote}
\begin{equation}
\langle \alpha | H^{eff} | \beta \rangle = 
\langle \alpha | T | \tilde{\beta} \rangle +
\langle \tilde{\alpha} | V + V {1 \over E-QH} QV | \tilde{\beta} \rangle
\end{equation}
where
\begin{equation}
| \tilde{\alpha} \rangle = {E \over E-QT} | \alpha \rangle
\end{equation}
with $| \alpha \rangle$ a HO Slater determinant.
Thus $| \alpha \rangle = | \tilde{\alpha} \rangle$ for Slater
determinants for which $\Lambda_\alpha \neq \Lambda_{SM}$.  Otherwise, apart
from the overall normalization, $| \tilde{\alpha} \rangle$ 
differs significantly from $|\alpha \rangle$ only in its 
large-$r$ behavior (Fig. 2).
We stress that Eq. (4) is equivalent to Eq. (1).
Finally we insert the bracketed Green's function expansion of Eq. (2)
into Eq. (4).  Thus $QV$ always appears between insertions of $QT$, 
summed to all orders.

The results are given by the solid lines in Figs.
1a and 1c.  The $\langle 36s|H^{eff}|36s \rangle$ matrix element
and total binding energy now converge rapidly.
In fact $\Lambda_{SM}$ can be lowered to $\sim 40$ while
keeping high momentum contributions perturbative: in third order
1 keV accuracy is maintained.
  
However, with further lowering of the SM scale, the convergence again
exhibits hints of deterioration -- with new symptoms.
For $\Lambda_{SM} = 30$, the third-order calculation reduces 
$\sim$ 10\% errors in the bare value of $\langle 1s | H^{eff} | 1s \rangle$
to $\sim 0.2$\%.  But an error in excess of 0.1\% -- corresponding
to 50 keV -- persists after 10 additional orders of perturbation.
Unlike the case discussed above, all matrix elements are 
affected, though the nonperturbative corrections to $s-s$ 
transitions are larger than those for $s-d$ and much larger than
those for $d-d$ matrix elements.  This is a signature of 
scattering at small $r$ through the hard core.  Numerically 
one can verify that the nonperturbative tail disappears if the
hard core in $V$ is removed.

An exact ET must yield the same result regardless of the 
choice of excluded space parameters $b$ and $\Lambda_{SM}$.
However it is possible that a judicious choice 
of these parameters might simplify the numerical difficulty of the ET.  While
$\Lambda_{SM}$ generally is limited by the size of one's
computer, $b$ can be varied freely.
A natural choice for $b$ is the value that minimizes the 0th order energy
(obtained by ignoring entirely the last term in Eq. (1) or Eq. (4)).
This corresponds to minimizing the bare Hamiltonian $\langle T+V \rangle_{SM}$ 
as a function of $b$ in Eq. (1). 
The closer the binding energy to the correct value, of course,
the smaller the contribution of the high-momentum corrections
due to scattering in the excluded space.
  
The dashed lines in Fig. 3 are the 0th-order energies for Eq. (1) as a function of $b$ 
for $\Lambda_{SM}$ = 6, 8, and 10.  As $\langle T+V \rangle_{SM}$ is
a variational estimate of the energy, the contour minima are upper bounds
to the exact result.  The 0th-order results fail to bind the 
deuteron; the minima are achieved for
$b \sim$ 0.79-0.83$f$.  (Note that $b_{eff}=\sqrt{2} b$ is the size scale
in the relative coordinate, $\vec{r_1}-\vec{r_2}$.)
This $b_{eff}$ is considerably smaller than the nuclear size: the SM
is doing its best to find a compromise between two needs,
resolving the hard core (a problem that becomes easier for 
small $b$) and reproducing the correct long-distance behavior
(a problem that becomes easier for large $b$ - a doubling of
$b$ roughly halves the number of high-momentum states that
must be included to calculate $T^{eff}$ to an equivalent accuracy).  The
resulting compromise $b$ addresses neither need well.
  
The solid lines in Fig. 3 are the corresponding results for
the 0th-order approximation to Eq. (4).  The minima shift to
$b \sim$ 0.4-0.5$f$, and the unperturbed results are very
accurate, with errors of 21, 36, and 52 keV for 10, 8, and
6 $\hbar \omega$ spaces -- an improvement of about a factor
of 100 in binding-energy accuracy over the corresponding dashed-line results.
This has been achieved without taking into account any 
effects of $QV$.  The interpretation is clear: once one has solved
the long-distance problem through the resummation of Eq. (4),
the nonperturbative effects (as we shall see below) of the
hard core can be absorbed almost entirely into the included space,
given an appropriate HO size scale $b \sim r_c$, where $r_c$  
is the hard-core radius.  This scale arises naturally out of the
minimization, because the short-range repulsion is such a
dominant feature of the potential.

The $4 \hbar \omega$ 0th-order energy minimum is not quite so
impressive: the included space is sufficiently restrictive that a
nonnegligible contribution from $QV$ remains even if $b$
is optimized.  Conversely, the minimum at 10$\hbar \omega$
is very flat, which means that there is a range of $b$ -- a
set of included spaces -- in which the effects of $QV$ remain very 
small.

Fig. 4 gives our most important results.  Fig. 4a 
illustrates the problem noted in the introduction, a perturbative
expansion of Eq. (1) converges very poorly.
Only at 10 $\hbar \omega$ (not shown) do the fluctuations begin to damp out, 
though they remain at the $\sim$ 100 keV level even at high 
orders.  In contrast, the expansion based on Eq. (4) (Fig. 4b) is 
highly perturbative.  The 1st-order correction $V (E-QT)^{-1} QV$
for the 6,8, and 10 $\hbar \omega$ calculations yields binding energies
accurate to $\lsim$3 keV.  We see that the 10 and 8 $\hbar \omega$ 
calculations are effectively identical at and beyond 2nd order;
the 10, 8, and 6 $\hbar \omega$ results merge at 4th
order.  Even the $4 \hbar \omega$ 0th-order result is quickly
corrected to an accuracy of 1.2 keV at 3rd order.

The accuracy of the unperturbed $10 \hbar \omega$ result of Eq. (4) is
comparable to that achieved in \cite{hs} by direct summation
of the high-momentum contribution to Eq. (1) to $\Lambda_\infty \sim 70$.  
Perhaps more important, however, is the promising result that
the remaining corrections associated with $QV$ are perturbative.
This has important implications for the complexity of $H^{eff}$
in heavier nuclei, suggesting that the number of nucleons in the 
excluded space that must be linked by $QV$ can be limited.
For example, $V (E-QT)^{-1} QV$ links only three nucleons.  Rapid
convergence in the perturbation expansion translates into a
hierarchy of three-body, four-body, etc, contributions to
$V^{eff}$ of rapidly diminishing importance.

So far the discussion has emphasized results and their interpretation
rather than technical aspects.  However, the technical aspects 
deserve discussion because many HO properties can be usefully
exploited:

\noindent
1) The operator $T$ is diagonal in a plane wave basis.  The
transformation into that basis is particularly simple for the HO,
as the Fourier transform of a HO is a HO. 

\noindent
2) The operator appearing in the Green's function, however, is $QT$,
which is more difficult to treat.  One can relate this Green's
function to the free Green's function $G_0 = {1 \over E-T}$, at
the cost of a matrix inversion in the included space.  
Using the included-space projector $P=1-Q$ we define operators acting
within the included space
\begin{equation}
\Gamma_0 = P G_0 P,~~\Gamma_1 = P G_0 V G_0 P,~~
\Gamma_2 = P G_0 V G_0 V G_0 P~~...
\end{equation}
One can then write the perturbative expansion of $H^{eff}$ as
\begin{eqnarray}
H^{eff} &=& (E-\Gamma_0^{-1}) + \Gamma_0^{-1} \Gamma_1 \Gamma_0^{-1}
+ \Gamma_0^{-1} (\Gamma_2 - \Gamma_1 \Gamma_0^{-1} \Gamma_1) \Gamma_0^{-1} \nonumber \\
&+& \Gamma_0^{-1} (\Gamma_3 - \Gamma_1 \Gamma_0^{-1} \Gamma_2 -
\Gamma_2 \Gamma_0^{-1} \Gamma_1 + \Gamma_1 \Gamma_0^{-1} \Gamma_1 \Gamma_0^{-1} \Gamma_1) \Gamma_0^{-1} + ...
\end{eqnarray}
where the terms correspond to the contributions from $T$,
$V$, $(QV)^2$, $(QV)^3$, ..., respectively.  Note that the matrix $\Gamma_0^{-1}$
differs from the matrix $P(E-T)P$ only in the entries where the
bra and ket belong to the last included ``shell''. 

\noindent
3) As $T$ is a sum over all relative momenta, an evaluation of
$G_0$ in an independent-particle basis will generate a 
product of two-particle correlations (with a dependence on each
relative partial wave).  In contrast, with Jacobi coordinates
$G_0$ is diagonal in momentum space
\begin{equation}
G_0 = {1 \over E - {1 \over 2M} (\dot{k}_1^2 + ... + \dot{k}_{A-1}^2)}
\end{equation}
where $\dot{k}_i$ is the momentum associated with the $i$th Jacobi coordinate.
Clearly the Jacobi basis is the simpler choice.  As noted before, if we 
work to some order in perturbation theory, then the number of 
nucleons interacting at one time in the excluded space will be
limited.  Thus, in general, there will be ``spectator" nucleons,
and the matrix element will have a spectator dependence 
corresponding to the overlap integral weighted by $\Gamma_0$.
The importance of such spectator dependence in a HO-based ET
was noted in \cite{zheng}.  The integrals over spectator nucleons
can be handled analytically and are reducible to finite sums over
gamma and incomplete gamma functions.

These properties allow one to evaluate the HO matrix elements of the operators
of Eq. (7), and thus the binding energy term by term in perturbation theory.
In the present case of the deuteron the relative-coordinate matrix element is
transformed into momentum space
\begin{eqnarray}
\langle n_f (\ell_f s) j m t m_t | \Gamma_{\alpha}(E) | n_i (\ell_i s) j m t m_t \rangle =
{(-1)^{n_f+n_i+\alpha+1} \over \hbar \omega_{eff}}
\left[ {4 (n_f-1)! (n_i-1)! \over \Gamma(n_f+\ell_f+1/2) \Gamma(n_i+\ell_i+1/2)} \right]^{1/2} \nonumber \\
\times \int_0^\infty d \rho_f {\rho_f^{\ell_f+2} e^{\rho_f^2/2} \over
\rho_f^2 + E_0} L_{n_f-1}^{\ell_f+1/2}(\rho_f^2)
\int_0^\infty d \rho_i {\rho_i^{\ell_i+2} e^{\rho_i^2/2} \over
\rho_i^2 + E_0} L_{n_i-1}^{\ell_i+1/2}(\rho_i^2) 
\langle \rho_f (\ell_f s) j m t m_t | V_{\alpha} | \rho_i (\ell_i s) j m t m_t \rangle
\end{eqnarray}
where $\omega_{eff} = \omega/2$, $L$ is a Laguerre 
polynomial, and $E_0$ is the dimensionless binding energy $|E|/\hbar \omega_{eff}$.
The momentum-space matrix element of the iterated potential is 
given by a simple recursion relation
\begin{eqnarray}
\langle \rho_f (\ell_f s) j m t m_t | V_{\alpha} | \rho_i (\ell_i s) j m t m_t \rangle &=&
\sum_{\tilde{\ell}} \int_0^\infty d \tilde{\rho}
{\tilde{\rho}^2 \over \tilde{\rho}^2 + E_0}
\langle \rho_f (\ell_f s) j m t m_t | V_1 | \tilde{\rho} (\tilde{\ell} s) j m t m_t \rangle \nonumber \\
&\times& \langle \tilde{\rho} (\tilde{\ell} s) j m t m_t | V_{\alpha-1} | \rho_i (\ell_i s) j m t m_t \rangle
\end{eqnarray}
where
$\langle \rho_f (\ell_f s) j m t m_t | V_1 | \tilde{\rho} (\tilde{\ell} s) j m t m_t \rangle$ is given by
\begin{equation}
{1 \over \hbar \omega_{eff}} \int_0^\infty dr r^2 \sqrt{{2 \over \pi}} j_{\ell_f}(\rho_f r)
\langle (\ell_f s) j m t m_t | V(r b_{eff}) | (\tilde{\ell} s) j m t m_t \rangle
\sqrt{{2 \over \pi}} j_{\tilde{\ell}}(\tilde{\rho} r)
\end{equation}
Note that all integration variables
are dimensionless.

In summary we have shown that conventional HO-basis effective interactions 
calculations involve both long- and short-distance nonperturbative
scattering.  The effects of such scattering can be absorbed into
the included space
by an appropriate summation of the relative kinetic energy to all orders,
followed by a tuning of the HO included space to absorb most of the hard-core
scattering.  This tuning results from minimizing the 0th-order
energy as a function of $b$.  In the test case of the deuteron, we find very
accurate 0th order results and very rapid convergence in further orders of
perturbation.  Existing Jacobi coordinate SM codes can treat light $1p$-shell
nuclei and three-body interactions \cite{petr}, which should allow
similar calculations to be done through 1st-order in the two-body
potential (0th order in the weaker three-body potential) in these
cases.

We thank Bob Wiringa for providing the $v18$ potential codes,
and Silas Beane and Martin Savage for helpful discussions.
This work was supported in part by US Department
of Energy grants DE-FG03-00ER41132 and DE-FC02-01ER41187.

\pagebreak

\pagebreak
  
\begin{figure}[b]
\epsfxsize=5.0in \epsfbox{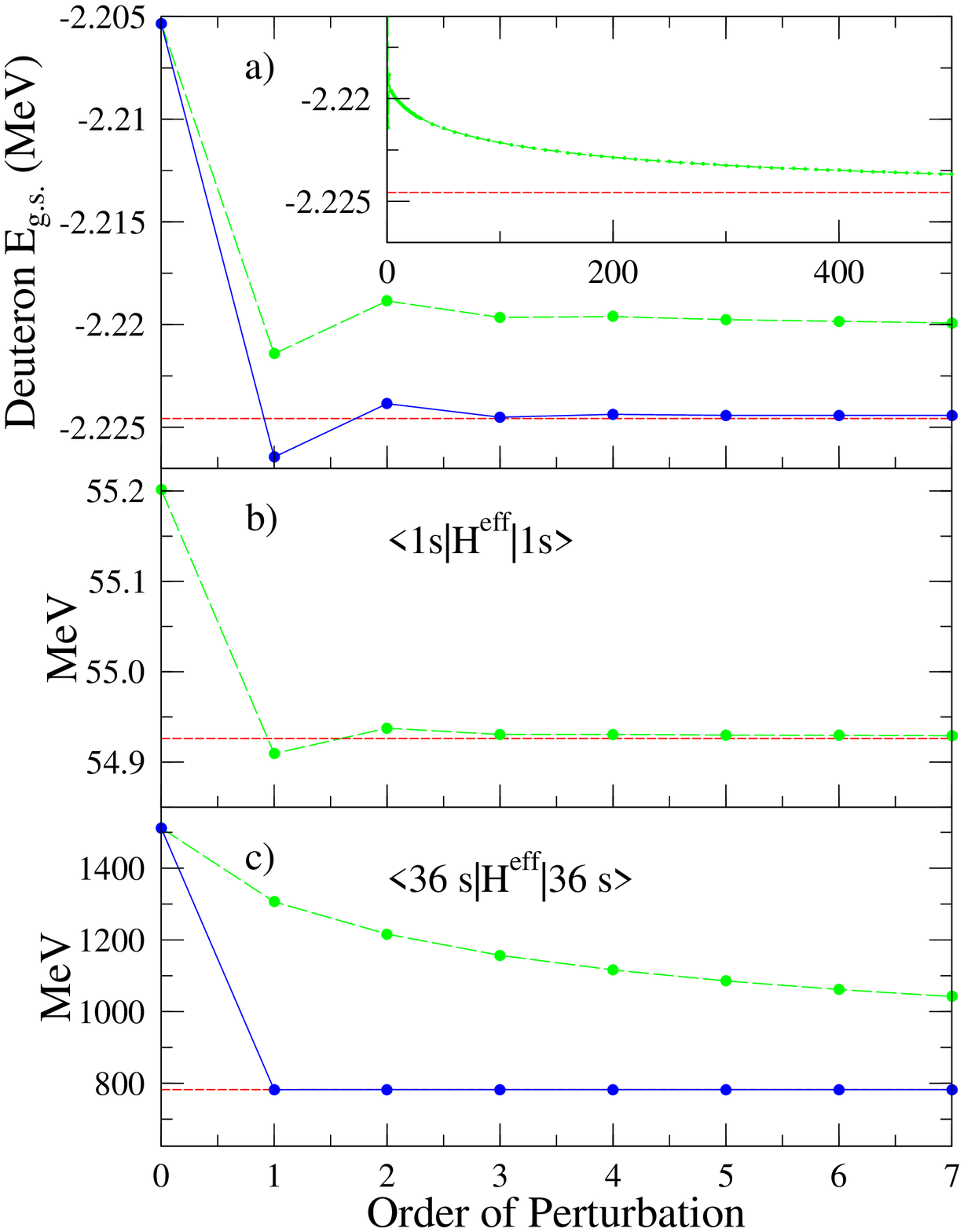}
\caption{The dash lines were obtained by inserting the Green's
function expansion of Eq. (2) into Eq. (1).  Very slow convergence
is found for the binding energy
(a) and for the effective matrix elements that involve included
states in the last ``shell.''
Other matrix elements (b) converge quickly.  The corresponding
results for Eq. (4) (solid lines) all converge rapidly.
The horizontal lines are the exact results.}
\end{figure}
  
\pagebreak
  
\begin{figure}[t]
\epsfxsize=5.3in \epsfbox{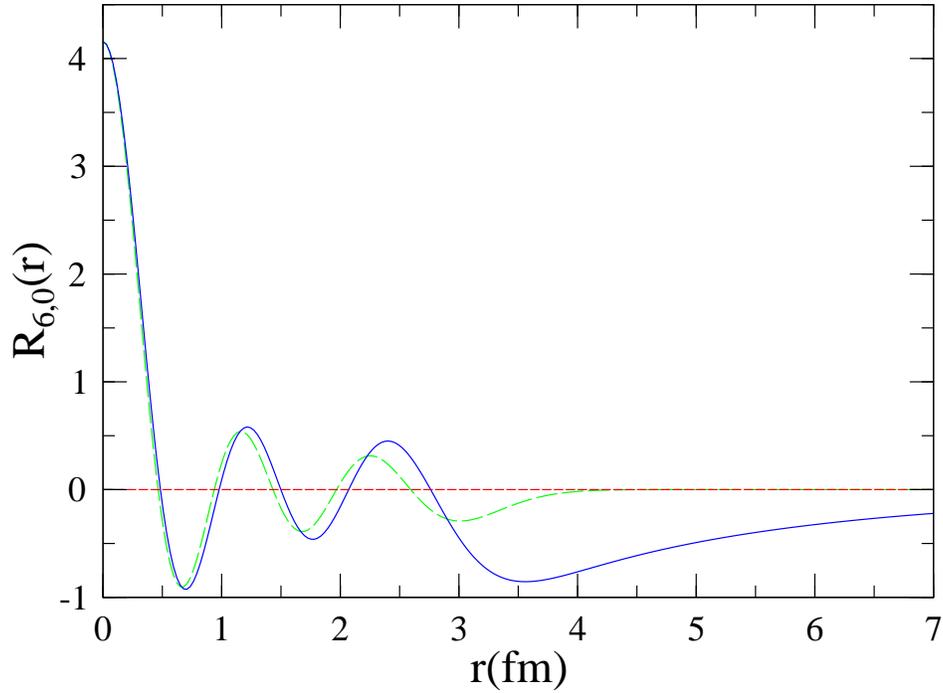}
\caption{For $\Lambda_{SM}=10$ the last included s-state has $n=6$.
The state $E/(E-QT) |n=6 \ell=0 \rangle$ (solid line) is compared
to its HO counterpart, $|n=6 \ell=0 \rangle$ (dashed line), showing that the
former is modified at large distances.
The modified wave function was multiplied by 1.934 to match the HO at $r=0$.}
\end{figure}

\pagebreak

\begin{figure}[t]
\epsfxsize=5.3in \epsfbox{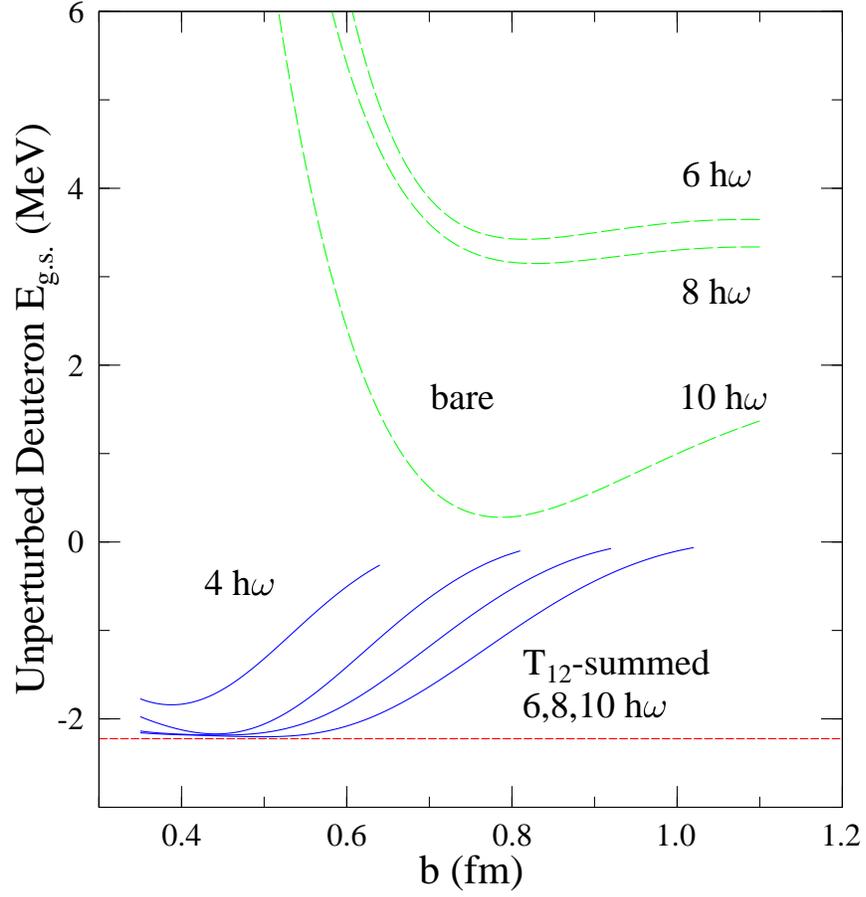}
\caption{The 0th order approximations to Eq. (1) (dashed lines)
and Eq. (4) (solid lines) as a function of $b$, showing the 
shifts in the minima and improved binding energies associated 
with the latter.}
\end{figure}

\pagebreak

\begin{figure}[t]
\epsfxsize=5.3in \epsfbox{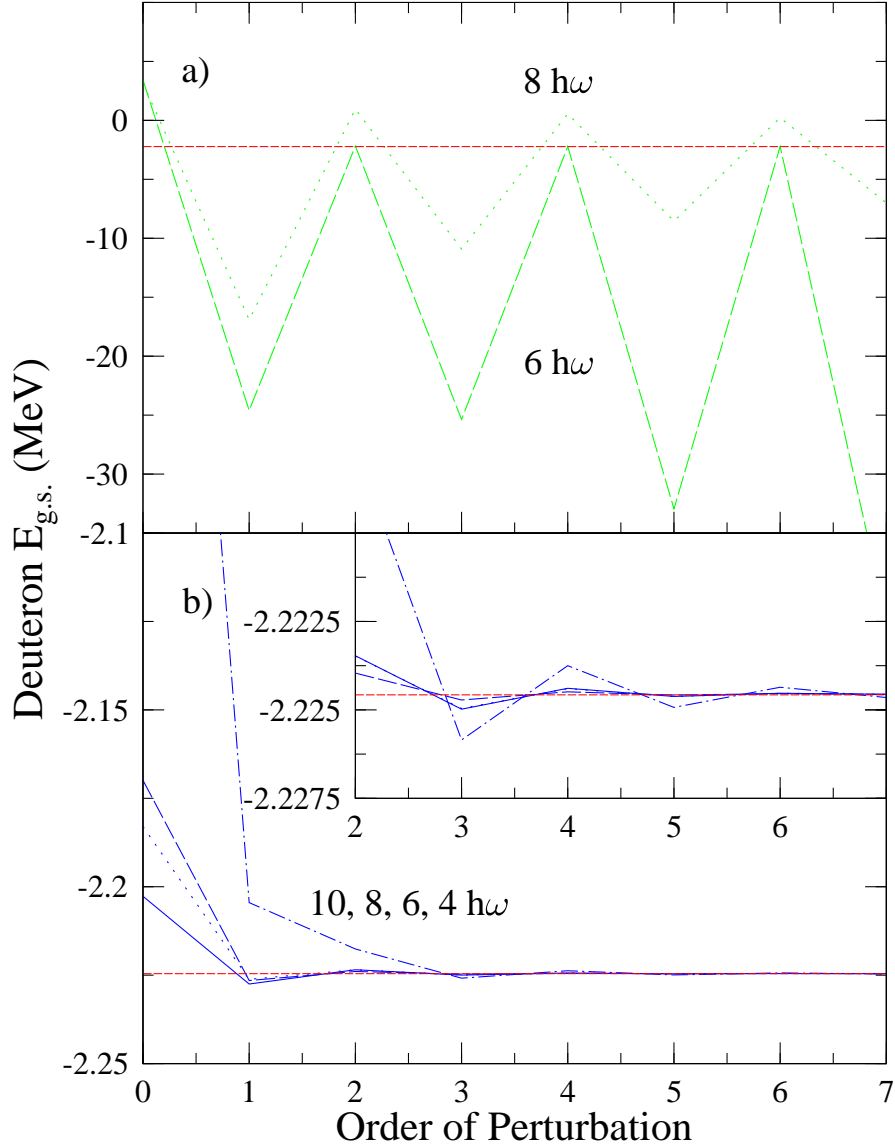}
\caption{Binding energies obtained by treating the excluded-space
contributions to Eq. (1) (a) and Eq. (4) (b) perturbatively,
via the Green's function expansion of Eq. (2).
The solid, dotted, dashed, and dot-dashed lines correspond to
10, 8, 6, and 4 $\hbar \omega$ included spaces, respectively.}
\end{figure}
\end{document}